\documentclass[12pt]{article}
\usepackage{fullpage}
\usepackage{setspace}
\usepackage{lscape}
\usepackage{amsmath}
\usepackage{amsthm}
\usepackage{acronym}
\usepackage{graphicx}
\usepackage{amssymb}
\usepackage{bm}
\usepackage{mathtools}
\usepackage{booktabs}
\usepackage{multirow}
\usepackage{hyperref}
\usepackage{setspace}
\usepackage{natbib}
\usepackage{color}
\usepackage{caption}
\usepackage{amsfonts}
\usepackage{bm}
\usepackage{mathrsfs}
\usepackage{mathptmx}
\usepackage{mathrsfs}
\usepackage{subcaption}
\usepackage{latexsym}
\usepackage{nicefrac}
\usepackage{verbatim}
\usepackage{lineno}
\usepackage{hyperref}
\hypersetup{
    colorlinks=true,
    linkcolor=black,
    citecolor=blue,
    urlcolor=blue
}

\everydisplay{\abovedisplayshortskip\belowdisplayshortskip}
\makeatletter \oddsidemargin  -.1in \evensidemargin -.1in

\theoremstyle{plain}

\title{Bayesian Bell regression model for fitting of overdispersed count data with application}
\author{Ameer Musa Imran Alhseeni\footnote{Email: amer.alhussani@gmail.com} \hspace{0.08cm} and Hossein Bevrani\footnote{Corresponding author. Email: bevrani@gmail.com}\\
\small {Department of Statistics, University of Tabriz, Tabriz, Iran}\\
}

\date{}

\begin{document}

\maketitle

\noindent{\bf {\em Abstract:}}
The Bell regression model (BRM) is a statistical model that is often used in the analysis of count data that exhibits overdispersion. In this study, we propose a Bayesian analysis of the BRM and offer a new perspective on its application. Specifically, we introduce a G-prior distribution for Bayesian inference in BRM, in addition to a flat-normal prior distribution. To compare the performance of the proposed prior distributions, we conduct a simulation study and demonstrate that the G-prior distribution provides superior estimation results for the BRM. Furthermore, we apply the methodology to real data and compare the BRM to the Poisson regression model using various model selection criteria. Our results provide valuable insights into the use of Bayesian methods for estimation and inference of the BRM, and highlight the importance of considering the choice of prior distribution in the analysis of count data.

\vskip 3mm
\noindent {\bf {\em {\color{blue} MSC2020 subject classifications:}}} 62F15, 62F25, 62J12.
\vskip 6mm

\section{Introduction}
The count regression models are valuable for understanding the relationships between predictor variables and count outcomes in various domains, and they offer a flexible and powerful framework for analyzing data with discrete and non-negative values. Some common applications of count data regression analysis include:
in biological and Genetic Studies to analyze data such as the number of genes, genetic mutations, or disease occurrences over time (\cite{Montesinos}), in epidemiology and public health  to analyze disease incidence, mortality rates, or the number of health events in a given population (\cite{Du}), in social sciences to model various phenomena such as the number of criminal offenses, births, or deaths in a specific area, in economic Research for analyzing economic events such as the number of business failures, patents, or the frequency of certain economic activities (\cite{Cameron}), insurance and actuarial science to model claim frequencies, policyholder behavior, and other related phenomena (\cite{Frees}) and finally, environmental studies to model ecological counts, such as the number of species in a habitat, wildlife populations, or environmental events (\cite{Lee}).

 Undoubtedly, the Poisson regression model is the primary choice for analyzing count data. This model, while widely used for analyzing count data, is subject to a significant limitation: it necessitates that the variance of the count variable is equal to its mean. This limitation can be particularly challenging when dealing with real-world count datasets, as overdispersion. Therefore, the model's applicability may be limited in such scenarios, prompting the need for alternative models. One  alternative is the Bell regression model (BRM) introduced by \cite{Castellares} and was highly acclaimed. BRMs have been wildly discussed in different situations, for example in the presence of multicollinearity (\cite{Majid, Ertan, Amin, Algamal1, Shewaa, Abduljabbar}), excess of zeros (\cite{Lemonte, Algamal2}), shrinkage strategies (\cite{Seifollahi1, Seifollahi2}).

In the field of statistical modeling, Bayesian inference has emerged as a powerful approach to data analysis. By combining prior knowledge with observed data, Bayesian methods enable the quantification of uncertainty and the estimation of model parameters. While the existing literature has mainly focused on frequentist approaches, the potential advantages of utilizing Bayesian methods in the estimation and inference of the BRM are apparent. This has motivated the authors to propose in this paper a Bayesian analysis of the BRM and provide a novel perspective on its application and extension. In this paper, a G-prior distribution is presented for Bayesian inference in BRM, along with a flat-normal prior distribution. Later, it is demonstrated through a simulation study that the G-prior distribution provides better estimations for the BRM.

The paper is structured as follows: Section \ref{sec2} provides a detailed explanation of the BRMs, which includes their prior development, posterior inference, and model selection criteria. In Section \ref{sec3}, a simulation study is carried out to compare the proposed prior distribution for BRMs. The methodology is then illustrated in Section \ref{sec4}, where BRM is compared to the Poisson regression model (PRM) using Bayesian inference. Finally, Section \ref{sec5} provides some concluding remarks.

\section{Model formulation and inference}\label{sec2}
\subsection{Bell regression model}
The discrete Bell distribution is introduced by \cite{Castellares} based on a series expansion due to \cite{Bell1, Bell2}. Its probability mass function is defined as:
\begin{equation}\label{bell:d}
f(y)= \dfrac{\theta^y e^{1-e^{\theta}} B_y}{y!}    \qquad y=0, 1, 2, \ldots \qquad \theta>0
\end{equation}
where $B_y=e^{-1}\sum_{k=0}^\infty k^n/k!$ indicates the Bell numbers.
The important characteristics of Bell distribution are given by

\begin{align*}
\mathbb{E}(Y)&= \theta e^{\theta}\\
\mathbb{V}(Y)&= \theta e^{\theta}(1+ \theta)= \mathbb{E}(Y) (1+ \theta)
\end{align*}
Since $ \theta>0 $, the variance value is greater than its mean which is known as overdispersion. Consequently, the regression model is defined based on Bell distribution will be suitable for count data with overdispersion. In regression contents, it is common to assume that the mean of the distribution depends on the vector of covariates. Let $\boldsymbol{y}=(y_1, y_2, \ldots, y_n)^T$ be a random sample from $Bell(\theta_i)$.
 We suppose  relate the $ \mu_i:=\mathbb{E}(Y_i)$ to $p$ covariates by the log link function, i. e.
\begin{equation}\label{modbell}
\log(\mu_i)= \boldsymbol{X}_i^T \boldsymbol{\beta} , \qquad i=1, 2, \ldots, n
\end{equation}
where $ \boldsymbol{\beta}=(\beta_0, \beta_1, \beta_2, \ldots, \beta_p)^T $ is the model parameter vector and  $\boldsymbol{X}_i=(x_{i1}, x_{i2}, \ldots, x_{ip})^T$ is the $i$th observation for $p$ model covariates. Due to \eqref{modbell}, the parameter of Bell distribution will be $\theta_i= W_0(\mu_i); \, i=1, 2, \ldots, n$ where $W_0(.)$ is the Lambert function. Hence, the model in \eqref{bell:d} can be reparameterized as:
\begin{equation}\label{ndist}
f(y_i)=\dfrac{ B_{y_i}}{y_i!} e^{1- e^{W_0(\mu_i)}}[W_0(\mu_i)]^{y_i}
\end{equation}

\subsection{Bayesian inference}
Consider $n$ observations of response-covariates pair as $\mathit{D}=\{(y_i, \boldsymbol{X}_i), \, i=1, 2, \ldots,n\}$. The likelihood function of the BRM is given as:
\begin{equation}\label{like_fun}
L(\mathit{D}\vert \boldsymbol{\beta})= \bigg(\prod_{i=1}^n \dfrac{ B_{y_i}}{y_i!}\bigg) \bigg(\prod_{i=1}^n [W_0(\mu_i)]^{y_i}\bigg) e^{n- \sum_{i=1}^n e^{W_0(\mu_i)}}
\end{equation}
where $\mu_i= \exp\big\{\boldsymbol{X}_i^T \boldsymbol{\beta}\big\}$.  Bayesian regression allows the incorporation of prior knowledge about the parameters, which is particularly useful when such information is available or when making use of expert knowledge. The inference is based on the posterior distribution of BRM, i.e.,
\begin{align}\label{post}
\pi(\boldsymbol{\beta} \vert \mathit{D})&= L(\mathit{D}\vert \boldsymbol{\beta}) \pi(\boldsymbol{\beta}) \nonumber\\
&=\bigg(\prod_{i=1}^n \dfrac{ B_{y_i}}{y_i!}\bigg) \bigg(\prod_{i=1}^n [W_0(\mu_i)]^{y_i}\bigg) e^{n- \sum_{i=1}^n e^{W_0(\mu_i)}} \times \pi(\boldsymbol{\beta})
\end{align}
Thus, we first need to specify the prior distribution.

\subsubsection{Specification of Priors}
First, we consider a common prior on $ \boldsymbol{\beta}=(\beta_0, \beta_1, \beta_2, \ldots, \beta_p)^T $ for our model, the flat-normal prior $N_p (\boldsymbol{0}, \tau^2 \boldsymbol{I}_p)$ where $\tau>0$. When we set a large value for $\tau^2$, the resulting flat prior becomes a diffuse prior. This diffuse prior assigns equal probability to all possible values of the regression coefficients. Unfortunately, this can lead to overestimating the magnitude of the regression coefficient and being overconfident about its sign. 
This is problematic because when it comes to regression coefficients (other than the intercept), we are typically more interested in knowing the magnitude and sign of the effect. Thus, we consider the second prior on model parameters based on the idea presented in \cite{Zhou}.

In a proposed regression model, if the mean $ \mu $ of $ y_i $ is assumed to be in the range of $(0,  \infty) $, and a subject matter expert has information on the marginal distribution of $ \mu $ characterized by $ IG(a_{\mu}, b_{\mu}) $, where $ a_{\mu} > 0 $ and $ b_{\mu} > 0 $ are known, the objective is to formulate a prior on $ \boldsymbol{\beta} $ that incorporates this prior information while adjusting for covariates. To achieve this, the G-prior proposed by \cite{Zellner} can be considered:
\begin{equation}\label{gprior}
 \boldsymbol{\beta} \sim N_p \bigg(M \boldsymbol{u}, gn (\boldsymbol{X}^T\boldsymbol{X})^{-1}\bigg)
\end{equation}
where $\boldsymbol{u}=(1, 0, \ldots,0)^T$, $M$ is a prior mean for the intercept and $g>0$ is a scalling constant.
Suppose that $\boldsymbol{X}_1, \boldsymbol{X}_2, \ldots, \boldsymbol{X}_n \stackrel{iid}{\sim}H(x)$ with the mean $\boldsymbol{A}$ and covariance matrix $\boldsymbol{\Sigma}$ such that $\mu(\boldsymbol{X}_i)= h^{-1}(\boldsymbol{X}_i^T\boldsymbol{\beta})$ and also assume that $\boldsymbol{X}_i$ and $ \boldsymbol{\beta} $ are independent. Therefore, we have
\begin{equation*}
\mathbb{E}(\boldsymbol{X}_i^T\boldsymbol{\beta})= \mathbb{E}\big(\mathbb{E}\big(\boldsymbol{X}_i^T\boldsymbol{\beta}\vert \boldsymbol{X}_i\big)\big)= \mathbb{E}\big(\boldsymbol{X}_i^T M \boldsymbol{u}\big)=M
\end{equation*}
and
\begin{align*}
\mathbb{V}(\boldsymbol{X}_i^T\boldsymbol{\beta})&= \mathbb{E}\big(\mathbb{V}\big(\boldsymbol{X}_i^T\boldsymbol{\beta}\vert \boldsymbol{X}_i\big)\big)
+\mathbb{V}\big(\mathbb{E}\big(\boldsymbol{X}_i^T\boldsymbol{\beta}\vert \boldsymbol{X}_i\big)\big)\\
&= \mathbb{E}\big(gn \boldsymbol{X}_i^T(\boldsymbol{X}^T\boldsymbol{X})^{-1}\boldsymbol{X}_i\big)+\mathbb{V}\big(M\big)\\
&= g. tr\bigg(n (\boldsymbol{X}^T\boldsymbol{X})^{-1} \big(\boldsymbol{\Sigma}+ \boldsymbol{A}\boldsymbol{A}^T\big)\bigg)
\end{align*}
Since $ n (\boldsymbol{X}^T\boldsymbol{X})^{-1} $ converges in probability to $ \big(\boldsymbol{\Sigma}+ \boldsymbol{A}\boldsymbol{A}^T \big)^{-1}$, thus
\begin{align*}
\mathbb{V}(\boldsymbol{X}_i^T\boldsymbol{\beta}) \stackrel{p}{\rightarrow} gp
\end{align*}
\cite{Hanson} found out for a various $H(.)$ considered in their simulations, $ \boldsymbol{X}_i^T\boldsymbol{\beta} $ is approximatly distributed as normal for any given value of $\boldsymbol{X}_i$.  Therefor, it is reasonable to consider that $\boldsymbol{X}_i^T\boldsymbol{\beta} \sim N(M, gp)$.
Using these results, we can select the value of $M$ and $g$ in G-prior distribution so that the induced distribution of $\mu(\boldsymbol{X}_i)= h^{-1}(\boldsymbol{X}_i^T\boldsymbol{\beta})$ matches the marginal prior distribution $\mu \sim IG(a_{\mu}, b_{\mu})$. By minimazing the Kullback-Leibler divergence
of these two distribution, we have $M = E\big(h(\mu)\big)$ and $g = \mathbb{V}\big(h(\mu)\big)/p$. In our regression model, we considered $h(.)$ as log link function, Hence,
\begin{equation}\label{gvalues}
M = \psi(a_{\mu})+\log(b_{\mu})  \qquad \text{and} \qquad g = \dfrac{1}{p}\psi^{(1)}(a_{\mu})
\end{equation}
where $ \psi(.) $ and $ \psi^{(1)}(.) $ are digamma and trigamma function, respectively.
When the values for $ a_{\mu} $ and $ b_{\mu} $ are not available, we use $a_{\mu}=b_{\mu} = 1$ as the defaults, yielding relatively weak prior information on the location of $\mu$.

 \subsubsection{MCMC algorithm}
Consider the posterior distribution by using flat-normal prior and G-prior distribution in \eqref{gprior} which is analytically intractable.
Monte Carlo Chain Markov (MCMC) simulation methods, such as the Gibbs sampler and Metropolis-Hastings algorithm, are utilized to obtain a sample from the posterior distribution (\cite{Gamerman}). 
To implement the Metropolis-Hastings algorithm, the following steps are taken:
\begin{enumerate}
\item Start with any point $\boldsymbol{\beta}^{(0)}$ and stage indicator $k=0$.
\item generate $\tilde{\boldsymbol{\beta}}$ according to the transitional kernel $K(\tilde{\boldsymbol{\beta}}, \boldsymbol{\beta}^{(k)})=N_p(\boldsymbol{\beta}^{(k)}, \tilde{\boldsymbol{\Sigma}})$ where $\tilde{\boldsymbol{\Sigma}} $ is a known symetric positive defined matrix.
\item Accept $\tilde{\boldsymbol{\beta}}$ as $ \boldsymbol{\beta}^{(k+1)} $ with the following probability
\begin{equation}
\min \bigg\{1, \dfrac{\pi (\tilde{\boldsymbol{\beta}}\vert \mathit{D}) }{\pi (\boldsymbol{\beta}^{(k)}\vert \mathit{D})} \bigg\}
\end{equation}
\item By increasing the stage indicator, repeat steps (1) to (3) until the process reaches a stationary distribution.
\end{enumerate}
The computational program is available upon request from the authors.

\subsection{Model selection Criteria}
There are multiple techniques available to compare and select the best-fitting model among several competing models for a given dataset. One commonly used technique in applied research is the conditional predictive ordinate (CPO) statistic. To learn more about the CPO and its applications in model selection, see \cite{Gelfand} and \cite{Eddy}.
Suppose $\mathit{D}$ the full data, $\mathit{D}_{(-i)}$ for $i=1, 2, \ldots,n$ denotes the data with the $i$th observation deleted and the posterior distribution based on  $\mathit{D}_{(-i)}$ is denoted by $\pi\big(\boldsymbol{\beta} \vert \mathit{D}_{(-i)}\big)$. For $i$th observation, $CPO_i$ is defined as follows
\begin{equation*}
CPO_i= \bigg[\int_{\boldsymbol{\beta}} \dfrac{\pi(\boldsymbol{\beta} \vert \mathit{D})}{f(y_i \vert\boldsymbol{\beta})}\, d\boldsymbol{\beta}\bigg]^{-1}.
\end{equation*}
The low CPO values indicate poor model fit, but a closed-form CPO is unavailable for the proposed model.
 However, a Monte Carlo estimate of the $ CPO_i $ can be obtained by using a single MCMC sample $ \{\boldsymbol{\beta}^{(1)}, \boldsymbol{\beta}^{(2)}, \ldots, \boldsymbol{\beta}^{(M)}\} $ from the posterior distribution $ \pi(\boldsymbol{\beta} \vert \mathit{D}) $. Thus, the $CPO_i$ can be approximated by
\begin{equation*}
\widehat{CPO}_i= \bigg[\sum_{k=1}^m \dfrac{1}{f(y_i \vert\boldsymbol{\beta}^{(k)})}\bigg]^{-1}.
\end{equation*}
The statistic for model comparison is the log-marginal pseudo-likelihood (LMPL) defined as follows:
\begin{equation}
LMPL= \sum_{i=1}^n \log (\widehat{CPO}_i)
\end{equation}
Therefore, the largest value of $LMPL$ indicates that the data is well-fitted by the model under consideration.
\\
The second criterion which is proposed by \cite{Spiegelhalter} is called the deviance information criterion (DIC). Based on MCMC samples, the DIC can be estimated as follows:
\begin{equation}
\widehat{DIC}= 2\bigg[- 2\dfrac{1}{M}\sum_{m=1}^M \log L(\mathit{D}\vert \boldsymbol{\beta}^{(m)}) +\log L(\mathit{D}\vert \bar{\boldsymbol{\beta}}) \bigg]
\end{equation}
where $ \bar{\boldsymbol{\beta}}$ is the mean of MCMC samples. The model with the lowest $\widehat{DIC}$ value is considered as the best-fitted model.
\\
The final  two criteria considered here are the expected Akaike information criterion (EAIC) by \cite{Brooks} and the expected Bayesian information criterion (EBIC) by \cite{Carlin}. These criteria can also be estimated using:
\begin{equation}
\widehat{EAIC}= -2\sum_{m=1}^M \log L(\mathit{D}\vert \boldsymbol{\beta}^{(m)}) +2p
\end{equation}
and
\begin{equation}
\widehat{EBIC} = - 2\sum_{m=1}^M \log L(\mathit{D}\vert \boldsymbol{\beta}^{(m)}) +p \log(n)
\end{equation}
where $p$ is the number of model parameter. The model that exhibits the lowest value of these criteria, similar to the DIC, is considered to be a better fit for the data.

\section{Simulation study}\label{sec3}
In this section, we conduct a simulation study to illustrate the implementation of the proposed regression methodology.
\\
The model we consider here is given by
\begin{equation}\label{md2}
\mu_i= \exp\big\{\beta_1x_{i1}+\beta_2 x_{i2}+\ldots+\beta_p x_{ip}\big\}, \qquad i=1, 2, \ldots, n
\end{equation}
Here, we consider the model with intercept, thus we set $x_{i1}=1$. The observations for the covariates, i. e.  $x_{i2}, \ldots, x_{ip}$, are generated from standard normal distribution. The real value of parameters in model \eqref{md2} are considered as follows:
\begin{equation}\label{realv}
\boldsymbol{\beta}=\big(0, -0.5, \underbrace{1, 1, \ldots, 1}_{p-2}\big)^T
\end{equation}
In summary, we obtained the response observations from $Bell\big(W_0(\mu_i)\big)$. In order to evaluate the effectiveness of our proposed method, we tested it with different sample sizes $n=50, 100, 200$ to represent low, medium, and high sample sizes, and with different numbers of covariates $p=3, 6$.
For the prior distributions, we consider flat-normal distribution with $\tau= 10^2$ and G-prior defined in \eqref{gprior} with hyperparameters obtained from \eqref{gvalues} by setting $a_{\mu}=b_{\mu} = 1$.

We generate two parallel independent MCMC run of size  $50,000$  for each posterior distribution  and discarded the $10,000$ first generated sample as burn-in
to eliminate the impact of initial values. For computing the posterior estimates, we used every $20$th sample to reduce the autocorrelations of the generated chains and yield better convergence results. 
The convergence of the MCMC chains was monitored using trace and autocorrelation function (ACF), Heidelberger-Welch,  and Gelman-Rubin convergence diagnostics. Additionally, we performed a small sensitivity study to evaluate the robustness of the model with respect to the choice of hyperparameters in the prior distributions, by testing different values of $\tau$ and $M$ for prior distributions. 

The summaries of the parameters in the posterior distribution exhibit minimal differences and do not impact the results presented in Tables \ref{Tab1} and \ref{Tab2}. These tables display the results of posterior inference for both prior distributions. These tables show the estimated values using the squared loss function, posterior standard deviations (PSD), and $95\%$ highest posterior density (HPD) intervals. 
The results show that the PSD and the width of the $95\%$ HPD interval decrease with an increase in the sample size. 
Moreover, the G-prior leads to estimations with lower PSD and narrower HPD intervals in comparison to the flat-normal prior.

Table \ref{Tab3} presents the mean squared errors (MSEs) and the mean absolute errors (MAEs) for the estimates found in Tables \ref{Tab1} and \ref{Tab2}.
This table demonstrates that MSE and MAE of G-prior is always lower than those of flat-normal prior except when $p=6$ and $n=200$.
For both prior distributions, increasing the number of covariates increase the MSE and MAE of estimators.

\vspace{0.5cm}
\begin{table}[!ht]
  \begin{center}
    \caption{Bayesian inference of model \eqref{md2} when $p=3$.}\label{Tab1}
 \footnotesize
\begin{tabular}{cccccccccccc}
\toprule
        & && \multicolumn{4}{c}{\text{G-prior}} && \multicolumn{4}{c}{\text{Flat normal}}\\
        \cmidrule{4-7} \cmidrule{9-12}
 & && & & \multicolumn{2}{c}{\text{$95\%$ HPD}} && & & \multicolumn{2}{c}{\text{$95\%$ HPD}} \\
        \cmidrule{6-7} \cmidrule{11-12}
 & \text{True value} &&\text{Estimate} & \text{PSD}& \text{Lower} & \text{Upper}  &&\text{Estimate} & \text{PSD}& \text{Lower} & \text{Upper}\\
   \hline   
$n=50$&&&&&&&&&&&\\[2pt]
$\beta_1$& 0 &&-0.0984&0.2022& -0.4795&  0.3073    &&-0.1088& 0.2135&-0.5245&  0.3017\\
$\beta_2$& -0.5 &&-0.4743&0.1830& -0.8319& -0.0946&&-0.4990& 0.1874&-0.8604& -0.1053\\  
$\beta_3$&  1   && 1.0369&0.1440& 0.7547&  1.3125  &&1.0629& 0.1482&0.7726&  1.3460\\
\hline
$n=100$&&&&&&&&&&&\\[2pt]
$\beta_1$& 0   &&-0.1609 &0.1430&-0.4254&  0.1273&&-0.1636&0.1465&-0.4449&  0.1175\\
$\beta_2$&-0.5&& -0.5825&0.1168&-0.8042& -0.3377&&-0.5903&0.1172&-0.8164& -0.3505\\
$\beta_3$&  1  &&1.0240  &0.0876&0.8479&  1.1900&&1.0329&0.0891&0.8569&  1.2067\\ 
\hline
$n=200$&&&&&&&&&&&\\[2pt]
$\beta_1$& 0    &&-0.1171&0.1051& -0.3133&  0.0911&&-0.1202&0.1081&-0.3203&  0.0926\\
$\beta_2$& -0.5&& -0.4977&0.0817&-0.6432& -0.3273&&-0.5007&0.0822&-0.6459& -0.3300\\
$\beta_3$&  1   &&1.0404&0.1067&0.8329&  1.2342&&1.0527&0.1091&0.8369&  1.2478\\
\bottomrule
\end{tabular}
\end{center}
\end{table}

\begin{table}[!ht]
  \begin{center}
    \caption{Bayesian inference of model \eqref{md2} when $p=6$.}\label{Tab2}
 \footnotesize
\begin{tabular}{cccccccccccc}
\toprule
        & && \multicolumn{4}{c}{\text{G-prior}} && \multicolumn{4}{c}{\text{Flat normal}}\\
        \cmidrule{4-7} \cmidrule{9-12}
 & && & & \multicolumn{2}{c}{\text{$95\%$ HPD}} && & & \multicolumn{2}{c}{\text{$95\%$ HPD}} \\
        \cmidrule{6-7} \cmidrule{11-12}
 & \text{True value} &&\text{Estimate} & \text{PSD}& \text{Lower} & \text{Upper}  &&\text{Estimate} & \text{PSD}& \text{Lower} & \text{Upper}\\
   \hline   
$n=50$&&&&&&&&&&&\\[2pt]
$\beta_1$&  0   &&0.0045&0.2046&-0.4025&  0.3899&&-0.0801&0.2286& -0.5406&  0.3511\\
$\beta_2$& -0.5&&-0.5517&0.1238&-0.8012& -0.3143&&-0.5742&0.1282&-0.8326& -0.3259\\  
$\beta_3$&  1   &&0.9397&0.1027&0.7257&  1.1282&&0.9798&0.1089&0.7535&  1.1779\\
$\beta_4$&  1   &&0.8249&0.1167&0.5838&  1.0548&&0.8750&0.1202&0.6326&  1.1170\\
$\beta_5$&  1   &&0.9214&0.1046&0.7168&  1.1318&&0.9283&0.1087&0.7099&  1.1381\\
$\beta_6$&  1   &&1.0821&0.1615&0.7607&  1.3963&&1.1515&0.1770&0.8025&  1.4978\\
\hline
$n=100$&&&&&&&&&&&\\[2pt]
$\beta_1$&  0   &&-0.1425&0.1471&-0.4346&  0.1352&&-0.2118&0.1588&-0.5130&  0.0938\\
$\beta_2$& -0.5&&-0.4448&0.0803&-0.6019& -0.2862&&-0.4491&0.0810&-0.6077& -0.2882\\  
$\beta_3$&  1   &&1.0378&0.0571&0.9199&  1.1507&&1.0550&0.0585&0.9329&  1.1725\\
$\beta_4$&  1   &&0.8337&0.1112&0.6046&  1.0405&&0.8784&0.1147&0.6420&  1.0970\\
$\beta_5$&  1   &&0.8680&0.0866&0.7029&  1.0404&&0.9007&0.0889&0.7356&  1.0770 \\
$\beta_6$&  1   &&1.0771&0.1061&0.8626&  1.2812&&1.1223&0.1132&0.8899&  1.3387\\
\hline
$n=200$&&&&&&&&&&&\\[2pt]
$\beta_1$&  0   &&-0.0152&0.0981&-0.2078& 0.1788&&-0.0406&0.1016&-0.2321&  0.1737\\
$\beta_2$& -0.5&&-0.4573&0.0542&-0.5611&-0.3551&&-0.4670&0.0546&-0.5710& -0.3645\\  
$\beta_3$&  1   &&0.8729&0.0674&0.7455&  1.0197&&0.8881&0.0684&0.7556&  1.0359\\
$\beta_4$&  1   &&0.9738&0.0577&0.8647&  1.0938&&0.9837&0.0596&0.8677&  1.1038\\
$\beta_5$&  1   &&1.0507&0.0552&0.9496&  1.1600&&1.0565&0.0562&0.9520&  1.1654\\
$\beta_6$&  1   &&0.9395&0.0658&0.8172&  1.0727&&0.9539&0.0669&0.8329&  1.0912\\
\bottomrule
\end{tabular}
\end{center}
\end{table}

\begin{table}[!ht]
  \begin{center}
    \caption{MSEs and MAEs for the estimated values reported in Tables \ref{Tab1} and \ref{Tab2}.}\label{Tab3}
     \footnotesize
\begin{tabular}{cccccccccccccc}
\toprule
 && \multicolumn{5}{c}{\text{MSE}} && \multicolumn{5}{c}{\text{MAE}}\\
        \cmidrule{3-7} \cmidrule{9-13}
&& \multicolumn{2}{c}{\text{G-prior}} && \multicolumn{2}{c}{\text{Flat normal}}&& \multicolumn{2}{c}{\text{G-prior}} && \multicolumn{2}{c}{\text{Flat normal}}\\
        \cmidrule{3-4} \cmidrule{6-7} \cmidrule{9-10} \cmidrule{12-13}
$n$&& \text{$p=3$} & \text{$p=6$}&&\text{$p=3$} & \text{$p=6$}&& \text{$p=3$} & \text{$p=6$}&&\text{$p=3$} & \text{$p=6$}\\
   \hline   
50  && 0.0036 & 0.0087  &&0.0052& 0.0093&&0.0537&0.0754&&0.0576&0.0871\\
100&& 0.0111& 0.0126 &&0.0120& 0.0150   &&0.0891&0.1018&&0.0956&0.1102\\
200&&  0.0051  & 0.0042  &&0.0057&0.0035&&0.0533&0.0537&&0.0579&0.0507\\
\bottomrule
\end{tabular}
\end{center}
\end{table}

\section{Application}\label{sec4}
In this section, we apply our methodology to a mine fracture dataset that was primarily analyzed by \cite{Myers}. 
The dataset includes four variables: the thickness of the inner burden in feet ($x_1$), the percentage of extraction from the previously mined lower seam ($x_2$), the height of the lower seam ($x_3$), and the time since the mine has been opened ($x_4$). The number of fractures in the mine is denoted by y, and there are 44 observations available for each variable.
 The authors consider two count regression models, the BRM and PRM, and they check whether the response variable follows these distributions using a chi-square test at a $95\%$ level of confidence. The chi-square test results, as shown in Table \ref{Tab4}, reveal that the Poisson distribution is not fitting well for this dataset, even though \cite{Myers} had previously applied the Poisson regression model to analyze this dataset.
\vspace{0.5cm}
\begin{table}[!ht]
  \begin{center}
    \caption{Goodness of fit test for the mine fracture dataset.}\label{Tab4}
     \footnotesize
\begin{tabular}{ccccc}
\toprule
Count & Observed & Bell & Poisson\\
\toprule
0  & 10 & 10.129 &6.474\\
1  & 7 &    7.446 &9.927\\
2  & 8 &    5.474 &7.611\\
3  & 8 &     3.353&3.890\\
4  & 4 &     1.849&1.491\\
$\geq5$& 7&1.748&0.606\\
\toprule
$\chi^2$ && 1.216 & 12.523\\
P-value   && 0.943&0.028\\
\bottomrule
\end{tabular}
\end{center}
\end{table}

The same as Simulation section, MCMC runs of size $50,000$ were generated for each posterior distribution of both BRM and PRM under the G-prior distribution. To reduce the autocorrelations of the generated chains and obtain better convergence results, every $20$th sample was used after discarding the first $10,000$ generated samples as burn-in. The posterior means, medians, PSDs, and $95\%$ HPD intervals for both regression models are presented in Table \ref{Tab5}. The results indicate that the BRM provides estimations with the lowest PSD and narrowest HPD intervals. Moreover, the HPD intervals of the BRMs indicate that $x_1$ and $x_3$ are not significant while only $x_2$ are significant based on the HPD intervals of PRM. 
To compare the BRM and PRM, the values of LMPL, DIC, EAIC, and EBIC criteria are computed and presented in Table \ref{Tab6}. Based on all the criteria, the BRM is identified as the most optimal model.
\vspace{0.5cm}
\begin{table}[!ht]
  \begin{center}
    \caption{Bayesian inference of models for the mine fracture dataset.}\label{Tab5}
\footnotesize
\begin{tabular}{cccccccc}
\toprule
& & & & & &\multicolumn{2}{c}{\text{$95\%$ HPD}} \\
\cmidrule{7-8}
\text{Model} &\text{Parameter} &  \text{Mean}& \text{Medain} &\text{PSD}&& \text{Lower} & \text{Upper}\\
   \hline
\multirow{5}{*}{\text{Bell}}&$\beta_0$&-3.5991&-3.5864  &1.0666&&-5.6350& -1.6295\\
&$\beta_1$&-0.0015&-0.0015  &0.0009&&-0.0032&  0.0002\\
&$\beta_2$&0.06310&0.0634  &0.0127&&0.0380&0.0879\\
&$\beta_3$&-0.0032&-0.0031  &0.0055&&-0.0148&0.0069\\
&$\beta_4$& -0.0323& -0.0316 &0.0164&&-0.0666&-0.0032\\
[2pt]
\hline
\multirow{5}{*}{\text{Poisson}} &$\beta_0$&-4.1262&-4.0515&1.4028&&-7.0078&-1.7307\\
&$\beta_1$& -0.0015&-0.0014&0.0011&&-0.0035& 0.0006\\
&$\beta_2$&0.0695&0.0689&0.0168&&0.0386&0.1031\\
&$\beta_3$& -0.0031&-0.0027&0.0074&&-0.0174&0.0116\\
&$\beta_4$&-0.0314&-0.0304&0.0227&&-0.0764&0.0117\\
\bottomrule
\end{tabular}
\end{center}
\end{table}

 \begin{table}[!ht]
  \begin{center}
    \caption{Bayesian Criteria for the fitted models to the mine fracture dataset.}\label{Tab6}
\begin{tabular}{cccccc}
\toprule
\text{Model} && \text{LMPL} & \text{DIC} & \text{EAIC} & \text{EBIC} \\
\hline
\text{Bell}      &&-72.4172&144.4801&149.3860&158.3070\\
\text{Poisson} &&-78.1424 &156.8615&161.5732&170.4942\\
\bottomrule
\end{tabular}
\end{center}
\end{table}

\section{Conclusions}\label{sec5}
We have prepared a paper describing a Bayesian technique for fitting BRMs. To the best of our knowledge, the potential advantages of utilizing Bayesian methods in the estimation and inference of the BRMs are apparent.
A G-prior distribution is proposed for Bayesian inference in BRMs, along with a flat-normal prior distribution.
The simulation results confirm the effectiveness of the proposed G-prior distribution in regression analysis. In the real data application, the BRM is compared to the Poisson regression model according to Bayesian inference, and the BRM outperforms the PRM according to several model selection criteria, including the LMPL, DIC, EAIC, and EBIC. The simulation study and application to real data results confirm the effectiveness of the proposed methodology in BRMs.


\end{document}